\documentclass[aps,prl,amsmath,amssymb,twocolumn,showpacs,superscriptaddress,final,floatfix.longbibliography]{revtex4-2}
\usepackage{natbib}
\usepackage{graphicx}	
\usepackage{xcolor}
\usepackage{color}	
\usepackage[normalem]{ulem}
\usepackage{hyperref}
\hypersetup{
  colorlinks,
  citecolor=blue,
  linkcolor=blue,
  urlcolor=blue}

\newcommand{\RFA}{RbFe$_2$As$_2$}
\newcommand{\KFA}{KFe$_2$As$_2$}
\newcommand{\CFA}{CsFe$_2$As$_2$}
\newcommand{\SRO}{Sr$_2$RuO$_4$}

\begin{document}

\title{Dispersion kinks from electronic correlations in an unconventional iron-based superconductor}


\author{M.-H. Chang}
\email{These authors contributed equally.}
\affiliation{Department of Physics, The Pennsylvania State University, University Park, Pennsylvania, 16802 USA}
\author{S. Backes}
\email{These authors contributed equally.}
\affiliation{RIKEN iTHEMS,  Wako, Saitama 351-0198, Japan; Center for Emergent Matter Science, RIKEN, Wako, Saitama 351-0198, Japan}
\author{D. Lu}
\affiliation{Stanford Synchrotron Radiation Lightsource, SLAC National Accelerator Laboratory, Menlo Park, California, 94025 USA}
\author{N. Gauthier}
\affiliation{Institut National de la Recherche Scientifique – Energie Matériaux Télécommunications, Varennes, QC J3X 1S2 Canada}
\author{M. Hashimoto}
\affiliation{Stanford Synchrotron Radiation Lightsource, SLAC National Accelerator Laboratory, Menlo Park, California, 94025 USA}
\author{G.-Y. Chen}
\affiliation{Center for Superconducting Physics and Materials, National Laboratory of Solid State Microstructures and Department of Physics, Nanjing University, Nanjing 210093, China}
\author{H.-H. Wen}
\affiliation{Center for Superconducting Physics and Materials, National Laboratory of Solid State Microstructures and Department of Physics, Nanjing University, Nanjing 210093, China}
\author{S.-K. Mo}
\affiliation{Advanced Light Source, Lawrence Berkeley National Laboratory, Berkeley, California, 94720 USA}
\author{R. Valent\'\i}
\email{valenti@itp.uni-frankfurt.de}
\affiliation{Institut f\"{u}r Theoretische Physik, Goethe-Universit\"{a}t Frankfurt, Max-von-Laue-Str.~1, 60438 Frankfurt am Main, Germany}
\author{H. Pfau}
\email{heike.pfau@psu.edu}
\affiliation{Department of Physics, The Pennsylvania State University, University Park, Pennsylvania, 16802 USA}

\date{\today}


\begin{abstract}

The attractive interaction in conventional BCS superconductors is provided by a bosonic mode. However, the pairing glue of most unconventional superconductors is unknown. The effect of electron-boson coupling is therefore extensively studied in these materials. A key signature are dispersion kinks that can be observed in the spectral function as abrupt changes in velocity and lifetime of quasiparticles. Here, we show the existence of two kinks in the unconventional iron-based superconductor \RFA~using angle-resolved photoemission spectroscopy (ARPES) and dynamical mean field theory (DMFT). In addition, we observe the formation of a Hubbard band multiplet due to the combination of Coulomb interaction and Hund's rule coupling in this multiorbital systems. We demonstrate that the two dispersion kinks are a consequence of these strong many-body interactions. This interpretation is in line with a growing number of theoretical predictions for kinks in various general models of correlated materials. Our results provide a unifying link between iron-based superconductors and different classes of correlated, unconventional superconductors such as cuprates and heavy-fermion materials.

\end{abstract}

\maketitle


\section{Introduction}

Unconventional and high-temperature superconductivity almost exclusively appears in materials with strong electronic correlations.
In order to determine if and to what extent correlations are a prerequisite, it is crucial to determine their effect on the electronic structure in the normal state out of which superconductivity emerges. 

Abrupt changes in the electron dispersion, called "kinks", are believed to hold important clues. Kinks can appear when electronic excitations couple to collective modes such as phonons, what suggests that they also act as the superconducting pairing mechanism. Therefore, the study of dispersion kinks attracted a tremendous amount of research in the past and many kinks have been observed experimentally \cite{Damascelli_2003,Sobota_2021,Iwasawa_2012,Tamai_2019,Wray_2008,Koitzsch_2009,Richard_2009,Kordyuk_2011,Malaeb_2014,Jang_2021}. However, their origin often remains controversial. In the context of correlated metals, it was predicted that strong interactions between electrons themselves can induce two kinks \cite{Byczuk_2007}. It is a feature in models of various strongly correlated systems \cite{Raas_2009,Deng_2013,Held_2013,Stadler_2021,Matsuyama_2017,Kugler2019,Hu_2020}, but observing both kinks experimentally has been challenging.

The spectral function of strongly correlated materials splits into two components: a coherent quasiparticle peak close to the Fermi energy $E_F$ and an incoherent part forming Hubbard bands at high energies. A kink separates the coherent from the incoherent part. Such kinks have been observed in the cuprates \cite{Ronning_2005,Meevasana_2007,Xie_2007,Valla_2007,graf_2007,Meevasana_2008} and ruthenates \cite{Iwasawa_2012} in the energy range between 300\,meV and 600\,meV. They are a direct fingerprint of strong electronic correlations \cite{Macridin_2007,Moritz_2009,Moritz_2010,Kim_2018,Wang_2020}. 

The models in Ref.~\onlinecite{Byczuk_2007,Raas_2009,Deng_2013,Held_2013,Stadler_2021,Matsuyama_2017,Kugler2019,Hu_2020} also predict a second kink that appears at low energies inside the quasiparticle band. The existence of this kink is less clear in other studies of the single-band Hubbard model \cite{Macridin_2007,Moritz_2009,Wang_2020}. The low-energy kink strongly renormalizes exactly those electronic excitations that condense into Cooper pairs in the superconducting state. Fermi liquid behavior only emerges below the kink energy \cite{Byczuk_2007,Deng_2013,Held_2013,Stadler_2021}. The kink has been suggested to be accompanied by the appearance of a collective spin excitation of similar energy \cite{Raas_2009}, which may be responsible for the superconducting pairing. A plethora of low energy kinks at a few tens of meV has experimentally been observed in unconventional superconductors such as cuprates \cite{Damascelli_2003,Sobota_2021}, ruthenates \cite{Tamai_2019} and iron-based superconductors \cite{Wray_2008,Koitzsch_2009,Richard_2009,Kordyuk_2011,Malaeb_2014}. However, their origin is often debated because they have an energy within the phonon as well as spin excitation spectrum. Only a few kinks could clearly be associated with electronic correlations \cite{Tamai_2019,Jang_2021}.

Here, we study the unconventional iron-based superconductor \RFA~with a combination of angle-resolved photoemission spectroscopy (ARPES) and fully charge self-consistent Density Functional Theory + dynamical mean-field theory (DFT+DMFT) calculations. We show that the strong electronic correlations in \RFA~separate the spectral function into a quasiparticle peak and a multiplet of Hubbard bands, which is the prerequisite for correlation-induced kinks. Indeed, we identify a low and a high-energy kink in ARPES and DMFT and argue that both originate from strong electronic correlations. \RFA~can therefore serve as a prototype material for the physics of correlation-induced kinks and it provides a unifying link between the various different classes of strongly correlated, unconventional superconductors such as cuprates, heavy fermions, nickelates, ruthenates, and iron-based superconductors.

\section{Results}


\begin{figure*}
\includegraphics[width=\textwidth]{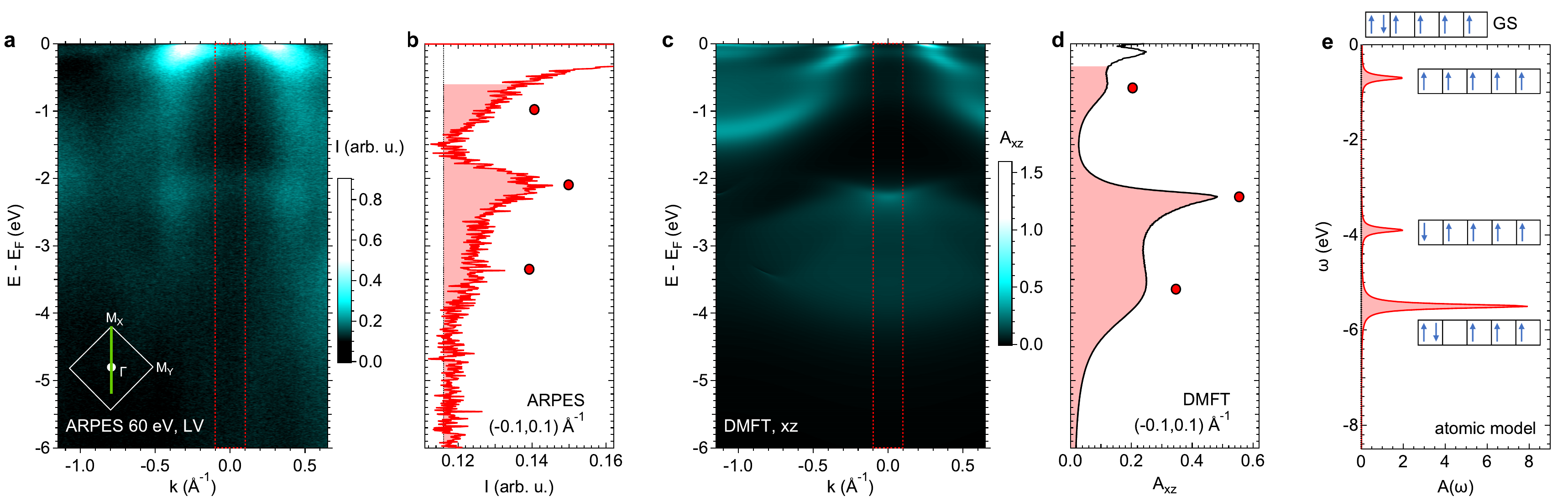}
\caption{
Hubbard bands in \RFA. {\bf{a}}: ARPES spectrum taken with 60\,eV, linear vertically (LV) polarized light and measured over a large binding energy window up to 6\,eV. Inset shows the momentum cut (green line) through the Brillioun zone (white square). {\bf{b}}: EDC extracted from the ARPES spectrum integrated over a momentum window of (-0.1,0.1)\,\AA (red dashed boxes), highlighting the incoherent spectral weight at high binding energy. {\bf{c}}: DMFT spectral function projected onto the $d_{xz}$ orbital in the same energy and momentum range as the ARPES data in {\bf{a}}. {\bf{d}}: EDC extracted from {\bf{c}} integrated over the same momentum window as {\bf{b}}. Red dots in {\bf{b,d}} mark the three peaks discussed in the main text. {\bf{e}}: Spectrum of electron removal states in an atomic model of five degenerate orbitals. Energy levels are broadened with a Gaussian for better visibility. The electronic configuration of the ground state (GS) is shown above the plot.
}
\label{Fig:hubbard}
\end{figure*}

\RFA~is a hole-doped, iron-based superconductor with a large Sommerfeld coefficient of 127\,$\mathrm{mJ/mol/K^2}$ \cite{Khim_2017,Zhang_2015}, large effective masses up to 25 times the bare electron mass \cite{Eilers_2016}, and low coherence and Fermi liquid temperatures below 100\,K \cite{Xiang_2016,Khim_2017,Wu_2016,Wiecki_2021}. These thermodynamic and transport measurements place \RFA~among the strongest correlated iron-based superconductors. A combination of Coulomb interaction and Hund's rule coupling is responsible for the correlated nature~\cite{Haule_2009,Medici2011,Medici2011PRL,Medici2014,Backes_2015,Stadler2019}.

The effect of correlations in the electronic structure of iron-based superconductors is most often described by renormalization factors that reduce the band width of the itinerant quasiparticles \cite{yi_2017}. At the same time, one expects a transfer of spectral weight from the coherent quasiparticle peak to the incoherent part of the spectrum. Therefore, we probe the electron-removal spectral function with ARPES across a large energy window down to 6eV as shown in Fig.\ref{Fig:hubbard}{a}. The ARPES spectrum was taken with a photon energy of 60\,eV, which probes a momentum close to $k_z=0$ \cite{Kong_2015}. The chosen light polarization predominantly leads to photoemission of electrons with $d_{xz}$ and $p_x$ orbital character along $k_x$. While the photon energy dependence of their relative photoemission cross sections is rather complex, energies beyond 50\,eV generally emphasize $d_{xz}$ spectral weight \cite{yeh_1985,Evtushinsky_2016,Watson_2017,Pfau_2021} (see methods section for further experimental details). The ARPES spectrum shows momentum-dependent incoherent spectral weight below -0.5\,eV. To highlight its energy dependence we plot an energy distribution curve (EDC) close to the Brillouin zone center in Fig.~\ref{Fig:hubbard}{b}. It shows a 3-peak structure (dots) that is energetically clearly separated from the sharp quasiparticle peak at low binding energies. 

Dynamical mean field theory combined with density functional theory calculations (DFT+DMFT) has proven to be a powerful method to describe and capture the correlated nature of metals such as iron-based superconductors~\cite{Haule_2009,aichhorn2010,yin2011,ferber2012lda,ferber2012fermi}. We therefore compare our ARPES results with the theoretical spectral function obtained from a DFT+DMFT calculation, projected onto the $d_{xz}$ orbital character, shown in Fig.~\ref{Fig:hubbard}{c}. We used established interaction parameters of $U_\mathrm{avg}=4$eV, $J_\mathrm{avg}=0.8$eV, representative for the iron-based pnictides \cite{Aichhorn2009,Roekeghem2016}  (see method section for further computational details). The EDC in Fig. \ref{Fig:hubbard}{d} shows the same three-peak structure (dots) of the incoherent spectral weight. Peak positions and relative intensities match the experimental observations. The region between -2\,eV and -5\,eV overlaps with the expected As $4p$ bands and simple $p$-$d$ hybridization may be responsible. However, the peak around -1\,eV is absent in DFT calculations (see Supplementary Note 1 \cite{supplement}).

A simple atomic model of five degenerate orbitals provides an intuitive understanding for the energy distribution of the incoherent spectral weight. In \RFA, nominally 5.5 electrons occupy the 10 Fe $3d$ states. However, a statistical analysis of the DFT+DMFT orbital configurations reveals that the average occupation is in fact close to 6 electrons. The ground state can therefore be approximated as a $d^6, S=\pm 4/2$ high spin configuration due to Hund's rule coupling \cite{Backes_2015}. The three possible electron removal states, which represent Hubbard bands, are depicted in Fig.~\ref{Fig:hubbard}{e}. Their energies were calculated using the same interaction parameters and assuming that the Fermi level is centered between the two lowest lying removal and addition states (see methods section for further details). Screening in a realistic lattice configuration is expected to decrease the energies of the transitions. The precise spectral weight distribution will be affected by additional effects that are not included in the atomic model, such as the specific $d$-orbital character, crystal field splittings, and hybridization with the As $p$-orbitals and resulting fluctuating valence state\cite{yin2011}.

The comparison with the atomic model demonstrates that the observed three incoherent peaks in ARPES and DFT+DMFT represent the multiplet of Hubbard bands. The first high-spin Hubbard satellite ($d^5, S=\pm 5/2$) around -1eV has previously been observed experimentally and theoretically in FeSe \cite{aichhorn2010,yin2011,Evtushinsky_2016,Watson_2017,Pfau_2021}. In addition, we expect a low-spin multiplet structure ($d^5, S=\pm 3/2$), which is located at higher energies due to the strong Hund's coupling effect. Arsenic $p$ orbitals appear at a similar energy range (see Supplementary Note 1 \cite{supplement}). Therefore, $p$-$d$ hybridization is expected to alter the intensity distribution of the $d_{xz}$ spectral function in this energy range. The extremely broad signatures in ARPES and DFT+DMFT between -2eV and -4eV in combination with the predictions from the atomic model demonstrate the existence of an additional low-spin multiplet structure ($d^5, S=\pm 3/2$) as would be expected for multiorbital systems \cite{Sroda2023}.

The clear separation of coherent quasiparticle peak and incoherent Hubbard bands places \RFA~into the realm of correlated materials that are predicted to show two kinks in the dispersion as described above.


\begin{figure*}
\includegraphics[width=\textwidth]{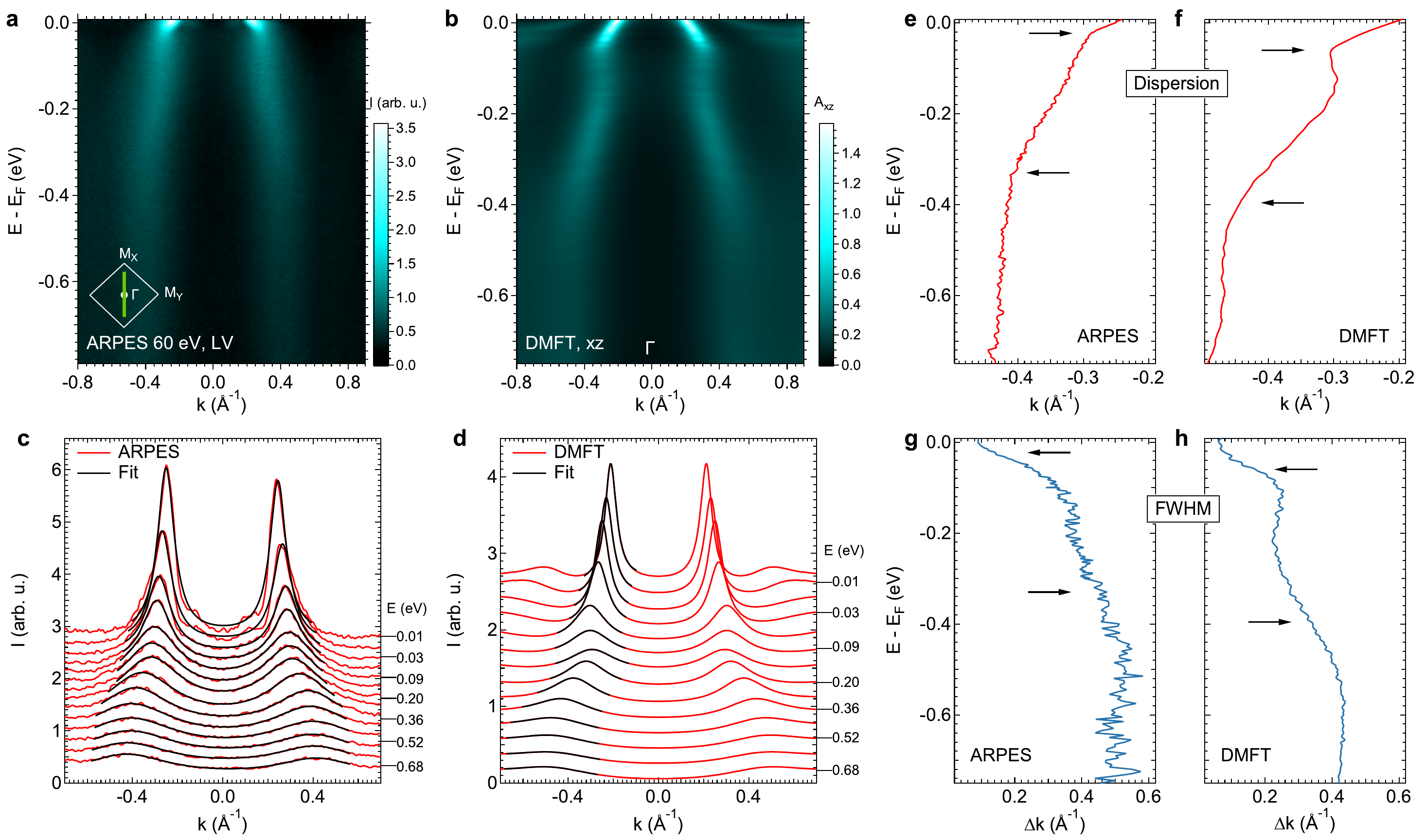}
\caption{
Spectral function of \RFA. {\bf{a}}: ARPES spectrum across the Brillouin zone center taken at 20 K with a photon energy of 60 eV. We use linear vertically (LV) polarized light, which highlights the $d_{xz}$ orbital. The spectrum was divided by a Fermi-Dirac function. Inset shows the momentum cut (green line) through the Brillioun zone (white square). {\bf{b}}: Spectral function along $k_x$ calculated by DMFT and projected onto the $d_{xz}$ orbital. {\bf{c}}: Selected momentum-distribution curves (MDCs) extracted from {\bf{a}} (red), which were fitted to a function that adds two Lorentzian peaks and a linear background (black). {\bf{d}}: same as {\bf{c}} for the spectral function shown in {\bf{b}}: with fits by single Lorentzian peaks. {\bf{e,g}}: Dispersion and FWHM extracted from the MDC fits. {\bf{f,h}}: Dispersion and FWHM extracted from MDC fits to the DMFT spectral function. The arrows in {\bf{e,f}}{\bf{,g,h}} mark the position of kinks. 
}
\label{Fig:disp}
\end{figure*}


A high-resolution ARPES spectrum across the energy scale of the expected kinks is presented in Fig.~\ref{Fig:disp}{a}. It clearly shows two dispersion kinks, which we verify by fitting of momentum distribution curves (MDCs) as exemplified in Fig.~\ref{Fig:disp}{c}. The extracted dispersion in Fig.~\ref{Fig:disp}{e} demonstrates the strong renormalization at the two kink energies of 23 meV and 330 meV.  The FWHM in Fig.~\ref{Fig:disp}g shows a step at each kink energy indicating a substantial increase in lifetime. Long-lived quasiparticles with a sharp spectral peak only develop at binding energies below the 23 meV kink. The flat dispersion in this energy range is consistent with the large effective mass from quantum oscillation measurements \cite{Eilers_2016}.

Typically, a self-energy is extracted from such an MDC analysis to compare with theoretical predictions. It requires the assumption of a bare-band dispersion, which would need to extend over more than 0.5eV in our analysis and certainly involves a-priori unknown non-linear contributions. To circumvent this ambiguity, we instead perform the same MDC analysis on the spectral function obtained from our DFT+DMFT calculations as we did for the ARPES data. The corresponding spectrum with a selection of MDCs, as well as the extracted dispersion and FWHM are shown in Fig.~\ref{Fig:disp}{b,d,f,h} respectively. The calculations reveal two kinks at almost the same energy scale and with very similar spectral appearance. The kinks represent the effect of the DMFT self-energy on the spectral function \cite{supplement}. Hence, they do not originate from simple hybridization effects of different Fe 3$d$ bands. A complementary self-energy analysis of the ARPES data can be found in the Supplementary Note 3 \cite{supplement}

The high energy kink at 330 meV is well beyond the highest energy in the phonon spectrum \cite{Richard_2018} and it is therefore not due to electron-phonon coupling. The kink initiates a high-energy section of steep dispersion with reduced intensity, which connects the coherent quasiparticle component of the spectral function with the first incoherent Hubbard band that is centered around 1\,eV binding energy. The experimental signatures are well reproduced by DMFT. The kink can therefore be associated with the high-energy kink predicted by general theoretical models of strongly correlated systems \cite{Byczuk_2007}. The steep high-energy dispersion is reminiscent of the "waterfall" structure seen in cuprate superconductors \cite{Ronning_2005,Meevasana_2007,Xie_2007}, where it develops from the spectral signature of the Mott insulating state \cite{Wang_2020}. Unusual high-energy renormalizations have also been observed in other iron-based superconductors \cite{Evtushinsky_2016,Watson_2017,Evtushinsky_2017,Pfau_2021}. Among them are both parent and electron-doped compounds, which are much further away from a Mott transition and therefore less correlated than the hole-doped \RFA. The same mechanism is likely responsible for these signatures and strong electronic correlations due to the interplay of Hubbard and Hund interactions therefore seem to play a crucial role for the vast majority of iron-based superconductors.

The kink energy of 23 meV lies within the phonon spectrum of \RFA \cite{Richard_2018}, which immediately suggests electron-phonon coupling as the origin. Such a kink is typically described with a Migdal-Eliashberg weak coupling mechanism. We can extract the corresponding coupling constant $\lambda = v_\mathrm{highE}/v_\mathrm{lowE}-1 = 3$ from the change in dispersion. However, $\lambda = 3$ is much larger than the weak coupling limit. It is also considerably larger than $0.5$ for kinks observed in various other FeSC \cite{Wray_2008,Koitzsch_2009,Richard_2009,Kordyuk_2011,Malaeb_2014}. 
Electron-phonon coupling from a standard weak-coupling mechanism is therefore unlikely to be responsible for the low-energy kink in \RFA.
A strong quasiparticle renormalization with an energy scale of approximately 20\,meV was also observed in FeTe, which is one of the strongest correlated iron-chalcogenides. This has been interpreted as polaron formation due to the spectral lineshape and the proximity of the kink to phonon energy scale \cite{liu_2013}. Another study suggested that spin-screening akin to Kondo coupling is responsible in line with an electronic origin \cite{Kim_2023}.

The comparison to DFT+DMFT calculations sheds further light on the origin of the low-energy kink in \RFA. Since the calculations do not include coupling to phonon excitations, all signatures are purely electronic in origin. The calculations show a similar kink at low energies. Apart from a difference in kink energy of approximately 35\,meV, all other experimental signatures shown in Fig.~\ref{Fig:disp} are very well reproduced by DMFT. In addition, the extracted Fermi velocities match almost perfectly (see Fig.~\ref{Fig:disp}{e,f} and Supplementary Note 5 for details \cite{supplement}). These similarities suggest that the experimentally observed low-energy kink originates predominantly from electronic interactions. 

Recent studies on FeSe demonstrated a cooperative enhancement of electron-electron and electron-phonon interactions and only taken together can they capture the experimental observations \cite{Mandal_2014,gerber_2017}. In particular, electronic correlations reduce the phonon mode frequencies in FeSe \cite{khanal_2020}. We speculate that such a cooperative mechanism influences the low-energy kink in \RFA~and reduces the kink energy compared to the DMFT calculations.

\begin{figure}
\includegraphics[width={\columnwidth}]{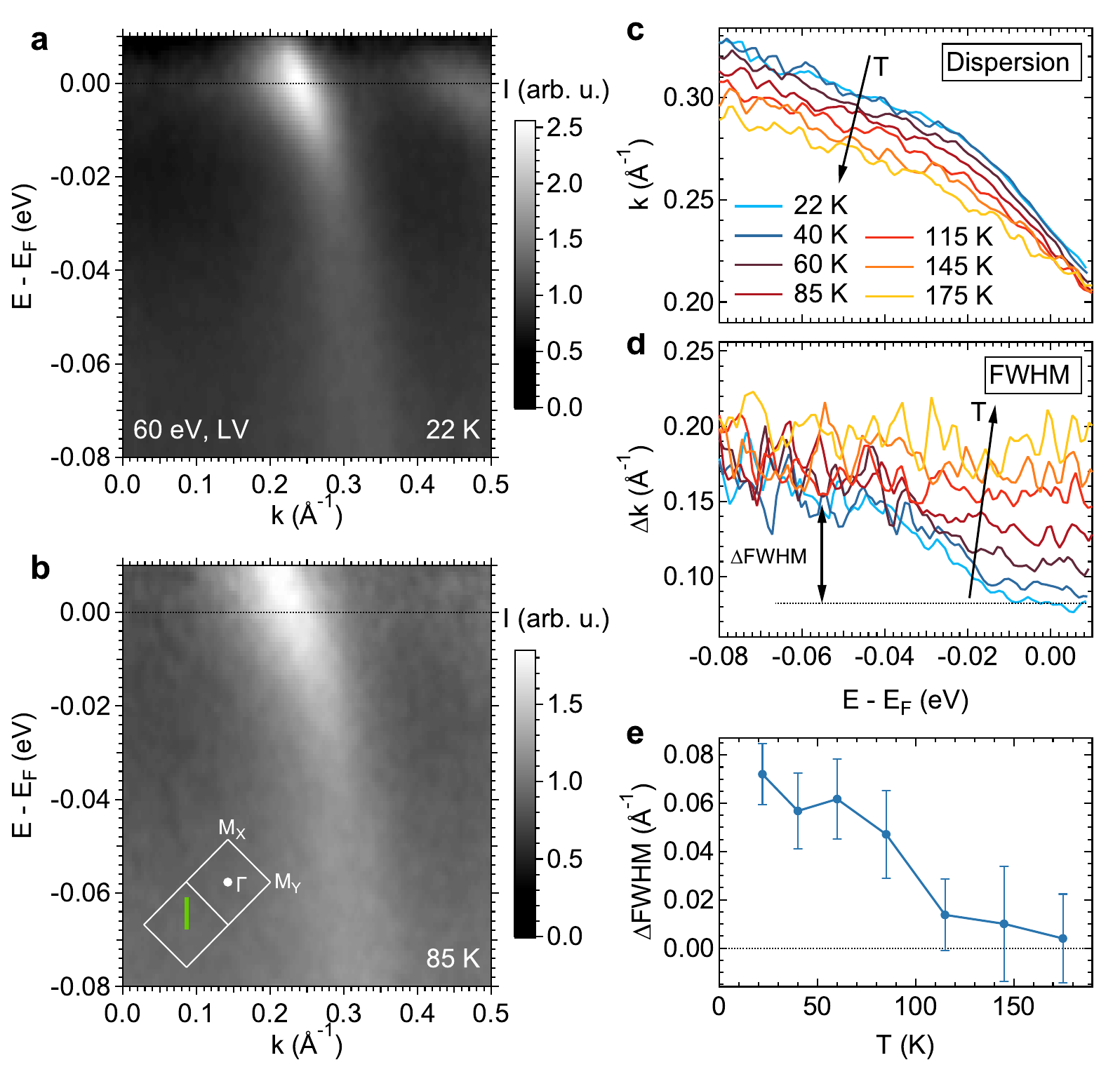}
\caption{
Temperature dependence of the low-energy kink. {\bf{a,b}}: ARPES spectra taken at 22\,K and 85\,K, respectively. We used linearly vertically (LV) polarized light with an energy of 60\,eV. The green line in the inset of {\bf{b}} sketches the momentum cut through the Brillouin zones. The dot indicates normal emission geometry. {\bf{b,c}}: We obtain the dispersion and FWHM from an MDC analysis of ARPES spectra between 22\,K and 175\,K. {\bf{e}} The step in the width $\Delta FWHM$ was determined from the difference in the average FWHM between (-70,-40)\,meV and (-10,8)\,meV as sketched in {\bf{d}}. Error bars represent the standard deviation.
}
\label{Fig:Tdep}
\end{figure}

Theoretical studies of the Hubbard, Kondo, and Hund-Hubbard model showed that the appearance of two kinks in correlated materials is related to a two-stage dynamics for establishing coherent quasiparticles \cite{Hu_2020,Stadler_2021}. It is therefore instructive to study the temperature dependence of the low-energy kink, which we show in Fig.~\ref{Fig:Tdep}. Note that the data were taken along a different momentum cut than those in Fig.~\ref{Fig:disp}{a}. Both the kink in the dispersion and the step in the FWHM disappear at higher temperatures. To quantify the behavior, we calculate the difference of the average FWHM below and above the kink energy ($\Delta$FWHM). Its temperature dependence shown in Fig.~\ref{Fig:Tdep}{e} indicates a crossover temperature around 70-100\,K, which coincides very well with the spin-coherence scale of 90\,K that was obtained from spin-susceptibility measurements \cite{Khim_2017}. Beyond this temperature, one expects a regime with damped quasiparticles, which agrees well with our experimentally observed increased width and decreased peak intensity. The analysis of the temperature dependence therefore supports our conclusion that the 23 meV kink is primarily electronic
in origin.
Very similar loss of renormalization and increased lifetime as function of temperature was also observed with ARPES and DMFT in the Hund's metal \SRO \cite{hunter_2023}. 

A low-energy correlation-induced kink implies the existence of spin fluctuations with a peak in the spin susceptibility close to the kink energy \cite{Raas_2009,Held_2013,Hu_2020,Stadler_2021}. It was suggested that the formation of the kink can be viewed as a coupling of electrons to these emergent internal bosonic modes \cite{Raas_2009}. One may speculate that such modes are also responsible for superconducting pairing. Unfortunately, neutron data are so far unavailable for \RFA. 
However, recent measurements on the sister compound \KFA~showed clear evidence for spin excitations below 20meV \cite{Lee_2011,Wang_2013} that develop into a resonance inside the superconducting state \cite{Shen_2020}. \KFA~has the same doping and very similar electronic and superconducting properties that evolve smoothly across the chemical pressure series from \KFA~over \RFA~to \CFA \cite{Backes_2015,Eilers_2016,Wu_2016,Xiang_2016,Wiecki_2021}. Evidence for nodes in the superconducting gap function was found experimentally \cite{Zhang_2015,Liu_2019,Okazaki_2012,Hashimoto_2010,Kim_2014,Cho_2016,Reid_2012,Dong_2010,Watanabe_2014}, but the superconducting pairing mechanism of these hole doped compounds is debated. Proposals range from spin-fluctuation mediated pairing \cite{suzuki_2011,maiti_2011,thomale_2011}, pairing due to strong Hund's electron-electron interaction \cite{Vafek_2017}, to pairing from orbital fluctuations \cite{Kontani_2010}.
Our discovery of a correlation-induced, strong low-energy kink on the same energy scale as spin fluctuations generates new perspectives for revealing the superconducting pairing mechanism in iron-based superconductors.


\section{Methods}

\subsection{ARPES Experiments}
Single crystals of \RFA~were grown from an FeAs flux \cite{Chu_2009}. ARPES measurements were performed at the SSRL synchrotron at beamline 5-2. Samples were glued onto copper sample holders and furnished with a ceramic top post using H20E epoxy. All gluing and curing steps were performed inside an argon glovebox. Samples were cleaved in-situ at a pressure of $3\cdot10^{-11}$ torr and below 30 K. We oriented the samples along the in-plane high-symmetry crystallographic directions using Fermi surface measurements. We used a photon energy of 60 eV and linear vertically polarized light. The slit of the ARPES analyzer was vertically oriented. A sketch of the geometry and calculations of the photoemission matrix elements \cite{goldberg_1978,gadzuk_1975,li_2024_matrix} can be found in the Supplementary Note 6 \cite{supplement}. The samples are generally very soft and tend to buckle after cleaving. In order to obtain data from a pristine flat part of the sample surface, we used a small beam spot size of approximately 50$\mu$m. The energy and angular resolution of the measurements is 16\,meV and 0.1$^{\circ}$, respectively. Data for Fig.\ref{Fig:hubbard} and \ref{Fig:disp} were obtained at 20\,K across the center of the first Brillouin zone. Data for Fig.~\ref{Fig:Tdep} were obtained in the second Brillouin zone at temperatures indicated in the figure. Additional supporting measurements have been performed at the MAESTRO beamline at the Advanced Light Source of the Lawrence Berkeley National Laboratory and at the Penn State Materials Research Institute.

\subsection{DFT+DMFT calculations}

For the fully charge self-consistent DFT+DMFT calculations we employ the implementation within the Wien2K package\cite{Wien2k_a,Wien2k_b} of the full-potential linear augmented plane wave (FLAPW) method in the generalized gradient approximation\cite{GGA}, using a $17\times 17 \times 17$ momentum grid discretization of the Brillouin zone. The Bloch wave functions were projected on the subspace of the correlated Fe $3d$ orbital\cite{Aichhorn2009,Ferber2014}, using a large window of $[-6,13.6]$eV, which includes the filled As $p$ states and higher energy unoccupied states, resulting in 30 bands on average for each momentum state. The impurity model of the Fe $3d$ subspace was solved with the continuous-time quantum Monte Carlo method in the hybridization expansion, using the segment picture\cite{Werner2006,Wallerberger2018,Bauer_2011}. The calculations were done at a temperature of $T=96K$, unless indicated otherwise. We used interaction parameters of $U_\mathrm{avg}=4$eV, $J_\mathrm{avg}=0.8$eV, representative for the iron-based pnictides\cite{Roekeghem2016}, in the definition of Slater integrals\cite{SlaterIntegrals}, and the nominal double counting correction\cite{nominalDC1,nominalDC2} with $N=5.5$ nominal filling. Real frequency data was obtained by performing stochastic analytical continuation\cite{beach2004} on the Matsubara self-energy, which was then used to calculate the momentum-resolved spectral function on real frequencies. Orbital occupations from DFT and DFT+DMFT, as well as projections of the spectral function onto other Fe $3d$ orbitals and the full spectral function summed over all states are presented in the Supplementary Note 2 \cite{supplement}.

\subsection{Atomic model}
For the atomic model we calculated the spectral function using exact diagonalization of a degenerate five orbital system with 6 electrons. We use the same interaction parameters $U_\mathrm{avg}, J_\mathrm{avg}$ as for DMFT. The ground state is a high spin configuration $d^6, S=\pm 4/2$. Within this model we can identify the following possible atomic multiplets: 3 electron removal states, which correspond to the three possible choices of removing an electron either from the doubly occupied orbital to result in a $d^5, S=\pm 5/2$, or  $S=\pm 3/2$ configuration. Or, removing an electron from one of the single occupied orbitals, resulting in a  $S=\pm 3/2$ configuration. There is only one possible electron addition state, resulting in $d^7, S=\pm 3/2$. Their removal (addition) state energies are at $(U-3J)/2$\,eV, $(U+5J)/2$, and $(U+9J)/2$\,eV ($-(U-3J)/2$\,eV), assuming that the Fermi level is centered between the two lowest lying removal and addition states.


\section{Acknowledgments}

 \begin{acknowledgments}
We are very grateful for valuable discussions with Jan von Delft, Rudi Hackl, Yu He, Patrick Kirchmann, Seung-Sup Lee, Brian Moritz and Zhi-Xun Shen. This work is supported by the U.S. Department of Energy, Office of Science, Office of Basic Energy Sciences, Materials Sciences and Engineering Division, under Award Number DE-SC0024135. RV acknowledges support from the Deutsche Forschungsgemeinschaft (DFG, German Research Foundation) -- CRC 1487, “Iron, upgraded!” -- project number 443703006. This research used resources of the Advanced Light Source, which is a DOE Office of Science User Facility under contract no. DE-AC02-05CH11231. Use of the Stanford Synchrotron Radiation Lightsource, SLAC National Accelerator Laboratory, is supported by the U.S. Department of Energy, Office of Science, Office of Basic Energy Sciences under Contract No. DE-AC02-76SF00515. The co-authors acknowledge use of the Penn State Materials Characterization Lab.
\end{acknowledgments}


\bibliography{main}

\end{document}


\title{Supplementary Information\\\vspace{5mm} \normalsize{"Dispersion kinks from electronic correlations in an unconventional iron-based superconductor"}\\\vspace{5mm}}


\author{M.-H. Chang}
\email{These authors contributed equally.}
\affiliation{Department of Physics, The Pennsylvania State University, University Park, Pennsylvania, 16802 USA}
\author{S. Backes}
\email{These authors contributed equally.}
\affiliation{RIKEN iTHEMS,  Wako, Saitama 351-0198, Japan; Center for Emergent Matter Science, RIKEN, Wako, Saitama 351-0198, Japan}
\author{D. Lu}
\affiliation{Stanford Synchrotron Radiation Lightsource, SLAC National Accelerator Laboratory, Menlo Park, California, 94025 USA}
\author{N. Gauthier}
\affiliation{Institut National de la Recherche Scientifique – Energie Matériaux Télécommunications, Varennes, QC J3X 1S2 Canada}
\author{M. Hashimoto}
\affiliation{Stanford Synchrotron Radiation Lightsource, SLAC National Accelerator Laboratory, Menlo Park, California, 94025 USA}
\author{G.-Y. Chen}
\affiliation{Center for Superconducting Physics and Materials, National Laboratory of Solid State Microstructures and Department of Physics, Nanjing University, Nanjing 210093, China}
\author{H.-H. Wen}
\affiliation{Center for Superconducting Physics and Materials, National Laboratory of Solid State Microstructures and Department of Physics, Nanjing University, Nanjing 210093, China}
\author{S.-K. Mo}
\affiliation{Advanced Light Source, Lawrence Berkeley National Laboratory, Berkeley, California, 94720 USA}
\author{R. Valent\'\i}
\email{valenti@itp.uni-frankfurt.de}
\affiliation{Institut f\"{u}r Theoretische Physik, Goethe-Universit\"{a}t Frankfurt, Max-von-Laue-Str.~1, 60438 Frankfurt am Main, Germany}
\author{H. Pfau}
\email{heike.pfau@psu.edu}
\affiliation{Department of Physics, The Pennsylvania State University, University Park, Pennsylvania, 16802 USA}

\date{\today}

\maketitle

\section{Supplementary Note 1: DFT bandstructure}
\begin{figure}
\includegraphics[width=\columnwidth]{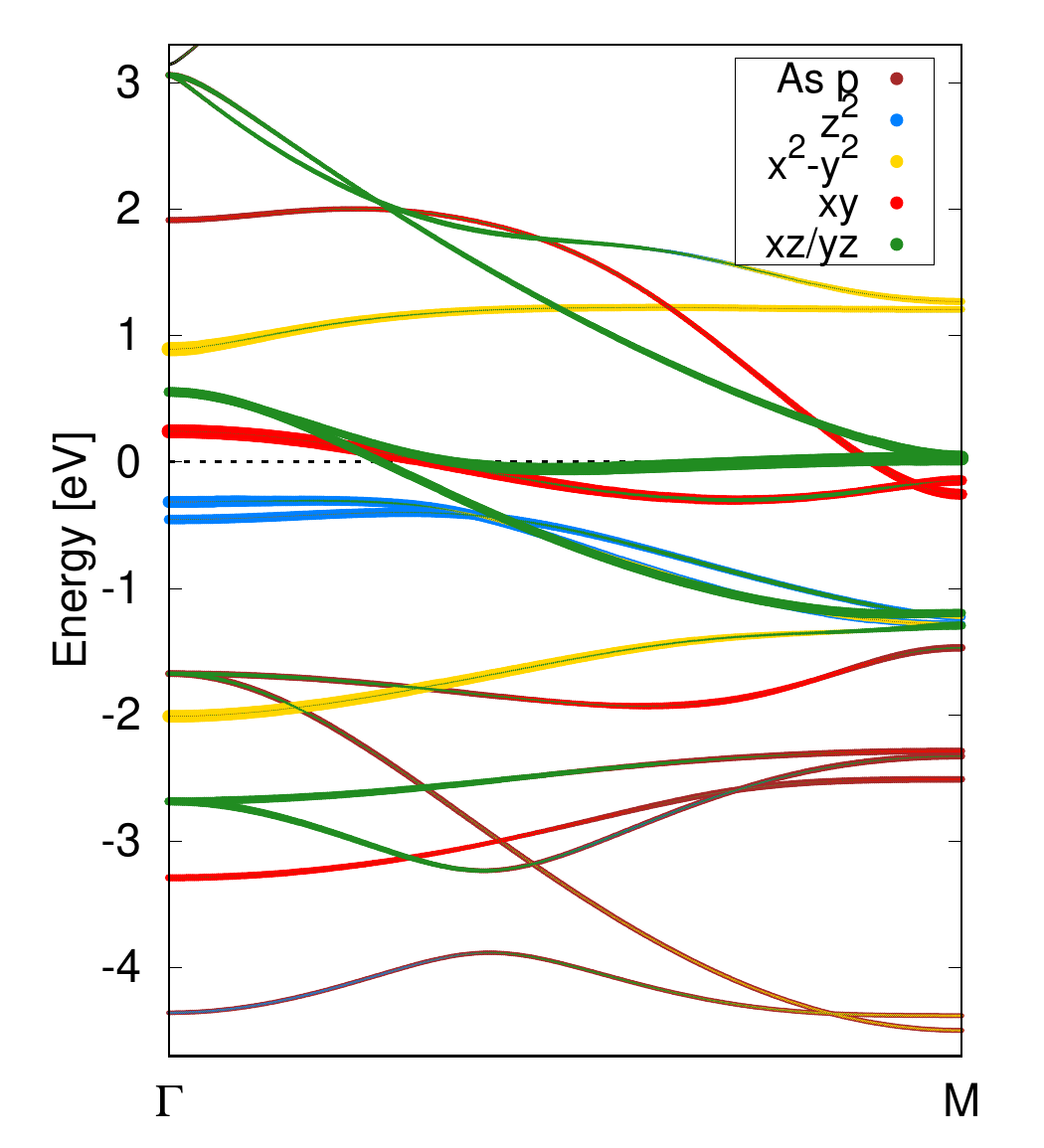}
\caption{
DFT bandstructure. The electronic band structure of {\RFA} along the $\Gamma-M$ path in the irreducible Brillouin zone as obtained from DFT (computational details are provided in the methods section).
Orbital character is indicated by the coloring and thickness of lines. \textit{As p} (brown) corresponds to the arsenic \textit{p} orbitals, and the remaining colors correspond to the Fe $3d$ orbitals.
}
\label{Fig:dftbands}
\end{figure}

In Supplementary Fig.\ref{Fig:dftbands} we show the electronic band structure of {\RFA} along the $\Gamma-M$ path as obtained from DFT, with colors indicating the projection on orbital character. As usual in the hole-doped 122 family, only hole-like Fermi surface pockets are found around $\Gamma$ and $M$. While the position of the As $p$ derived bands are in reasonable agreement with the experimental ARPES spectrum and the DFT+DMFT calculation, the Fe $d$ bandwidth is about 6-8 times larger. As described in the main text, there is no feature visible at $-1$~eV close to and around $\Gamma$, contrary to experiment and the DFT+DMFT calculation, indicating the existence of a correlation induced Hubbard-like satellite in {\RFA}.

\section{Supplementary Note 2: DMFT spectral function for all $3d$ orbitals}

\begin{figure*}
\includegraphics[width=\textwidth]{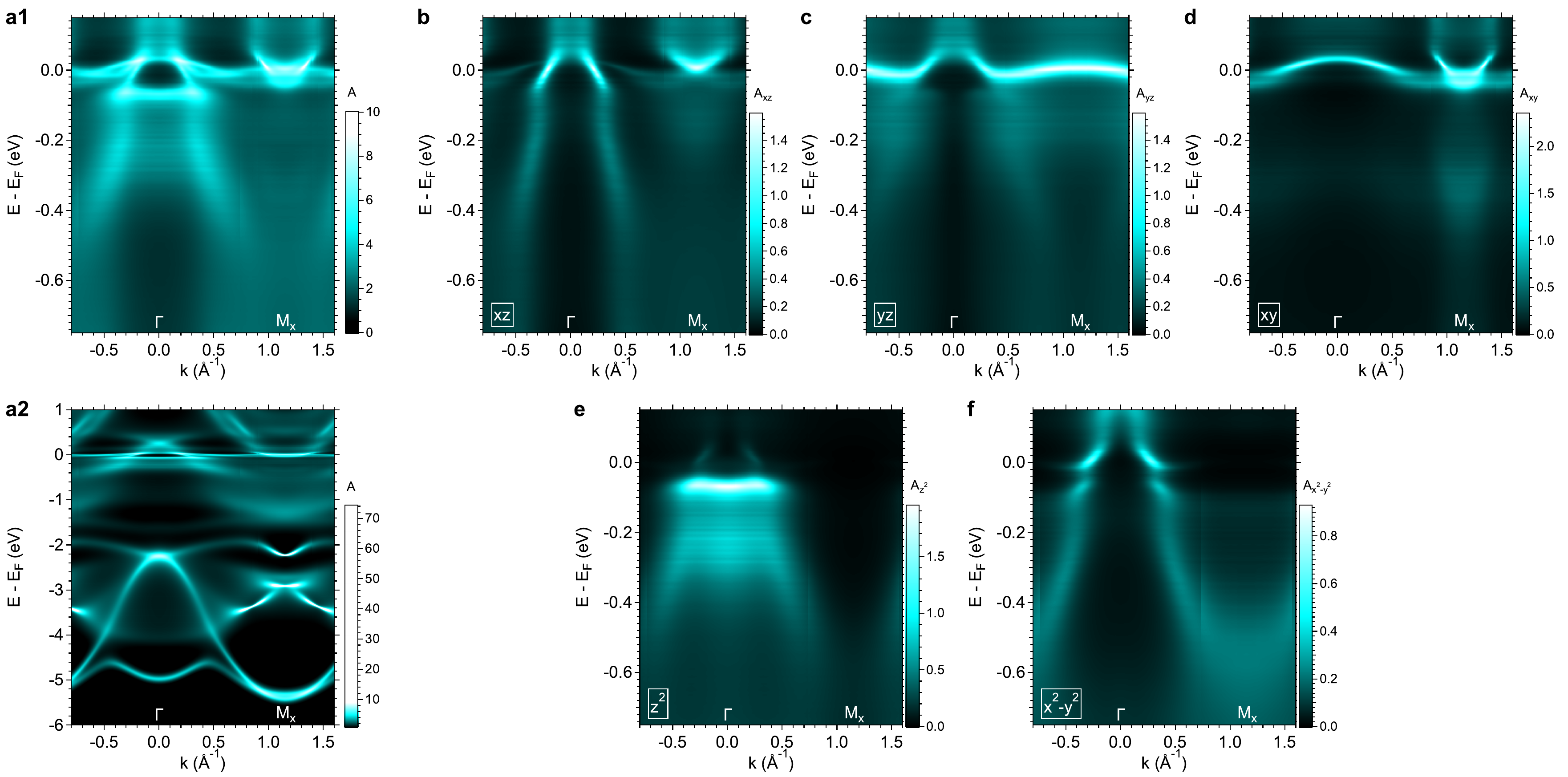}
\caption{
Spectral function from DFT+DMFT. (a1,a2) Spectral function summed over all states, including the As $p$ orbitals, in a small and a large energy window, respectively. (b-f) spectral function projected onto the five different Fe $3d$ orbitals as labeled in each plot.
}
\label{Fig:DMFT}
\end{figure*}

Supplementary Figure \ref{Fig:DMFT} shows the spectral function from DFT+DMFT projected onto the different Fe $3d$ orbitals (Fig.~\ref{Fig:DMFT}{b} to {f}) and the total spectral function summed over all states (Fig.~\ref{Fig:DMFT}{a1,a2}). Signatures of the kinks are visible in other orbitals as well. The particular dispersion of the $d_{xz}$ hole band in combination with favorable photoemission matrix elements provides the most favorable conditions to observe the kinks and analyse them in detail for both DMFT and ARPES.
In the total spectral function the As $p$ states appear as very sharp bands (Supplementary Fig.~\ref{Fig:DMFT}{a2}), since they are weakly correlated and only experience correlations via the hybridization with the Fe $3d$ states in the DFT+DMFT approach.

\section{Supplementary Note 3: DMFT and ARPES Self-energy}

In Supplementary Fig.~\ref{Fig:selfenergy}{a} and {b} we show the local part of the electronic self-energy for the Fe $d_{xz}$ orbital, obtained from the DFT+DMFT calculation and analytic continuation to the real frequency axis. The data appears noisy due to the stochastic analytic continuation procedure (see methods section). 
We obtain an effective quasiparticle weight $Z=\left( 1-\partial_{\omega} \mathrm{Re} \Sigma(\omega=0) \right)^{-1}\approx 0.167$,  and scattering rate $-Z \mathrm{Im}\Sigma(\omega=0) \approx 0.0167$~eV at a temperature $T=96$~K, indicative of strongly correlated bad metal behavior.
The real part shows the typical inverted slope and kink\cite{Mravlje2011,Kugler2019}.

Extracting the self-energy from ARPES measurements requires a bare band dispersion, which is a priory unknown. We use here the simplest approximation of a linear band as shown in Supplementary Fig.~\ref{Fig:selfenergy}{c}. The real part of the self-energy can then be approximated by the difference of the binding energy between ARPES and the bare band $ \mathrm{Re} \Sigma \approx E_k - E_\mathrm{lin}$. The imaginary part can be derived from the FWHM and the velocity of the bare band $\mathrm{Im} \Sigma \approx \mathrm{FWHM}\, v_\mathrm{lin}$. The overall shape and magnitude of the ARPES self-energy (Supplementary Fig.~\ref{Fig:selfenergy}{d,e}) matches the results from DMFT quite well. In particular, the signatures of the kinks are clearly visible.

\begin{figure*}
\includegraphics[width=0.8\textwidth]{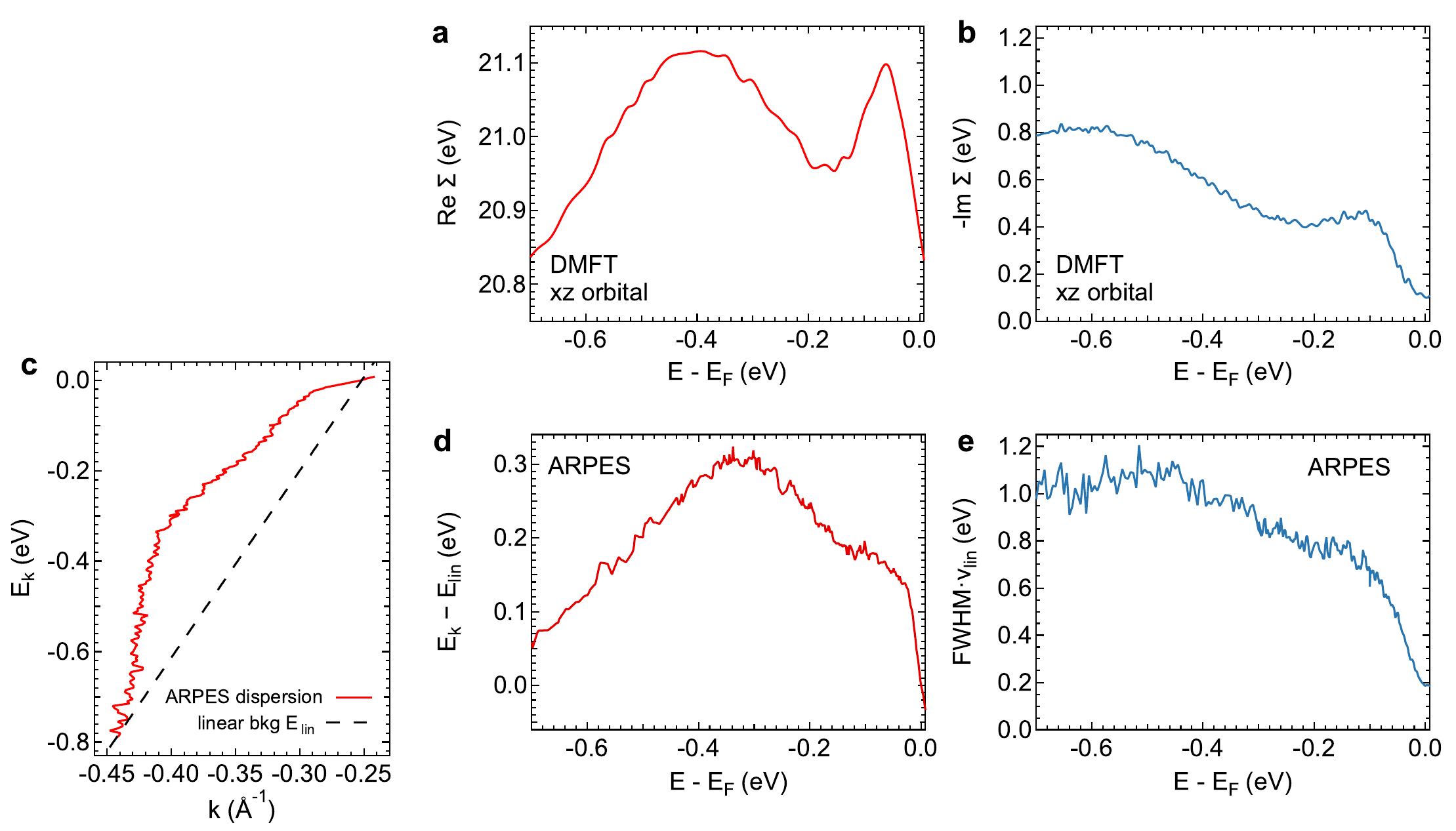}
\caption{
Self-energy. {\bf{a}}: Real part and {\bf{b}}: imaginary part of the DMFT self-energy of the $d_{xz}$ orbital. 
{\bf{c}}: ARPES dispersion reproduced from Fig.~2 of the main text (solid line) and linear approximation of a bare band (dashed line). {\bf{d,e}} Approximation of the real part of the self-energy $ \mathrm{Re} \Sigma \approx E_k - E_\mathrm{lin}$ and of the imaginary part of the self-energy $\mathrm{Im} \Sigma \approx \mathrm{FWHM}\, v_\mathrm{lin}$ using the bare band dispersion in {\bf{c}}.
}
\label{Fig:selfenergy}
\end{figure*}


\section{Supplementary Note 4: Orbital occupations}

\begin{table}[h]
    \centering
    \begin{tabular}{|c|c|c|}
    \hline
        orbital & DFT & DMFT \\
    \hline
       $z^2$     & 0.78 & 0.66 \\
       $x^2-y^2$ & 0.71 & 0.62 \\
       $xy$      & 0.81 & 0.59 \\
       $xz/yz$   & 0.79 & 0.59 \\
    \hline
    \end{tabular}
    \caption{Fe $3d$ orbital occupation numbers as obtained from DMFT}
    \label{tab:orbital_fillings}
\end{table}

In Table~\ref{tab:orbital_fillings} we show the Fe $3d$ orbital occupation numbers as obtained from DFT and DMFT. The average filling in DFT is close to 7.7 electrons in 5 orbitals, i.e. larger than the nominal 5.5 electrons filling due to hybridization with As $p$ orbitals. Introducing electronic correlations on the Fe $3d$ orbitals pushes a significant amount of charge from the Fe $3d$ states to the As $p$ states, resulting in an average filling close to 6 electrons in 5 orbitals.


\section{Supplementary Note 5: Fermi velocity}

We show a close-up view of the low energy dispersion obtained from MDC fits of the ARPES spectrum and DFT+DMFT calculations in Supplementary Figure \ref{Fig:lowE_kink}. We fit the dispersion within a the low-energy region with a linear function and obtain a Fermi velocity of $v=0.62$\,eV\AA~and $v=0.57$\,eV\AA~from ARPES and DMFT, respectively.

\begin{figure}
\includegraphics[width={\columnwidth}]{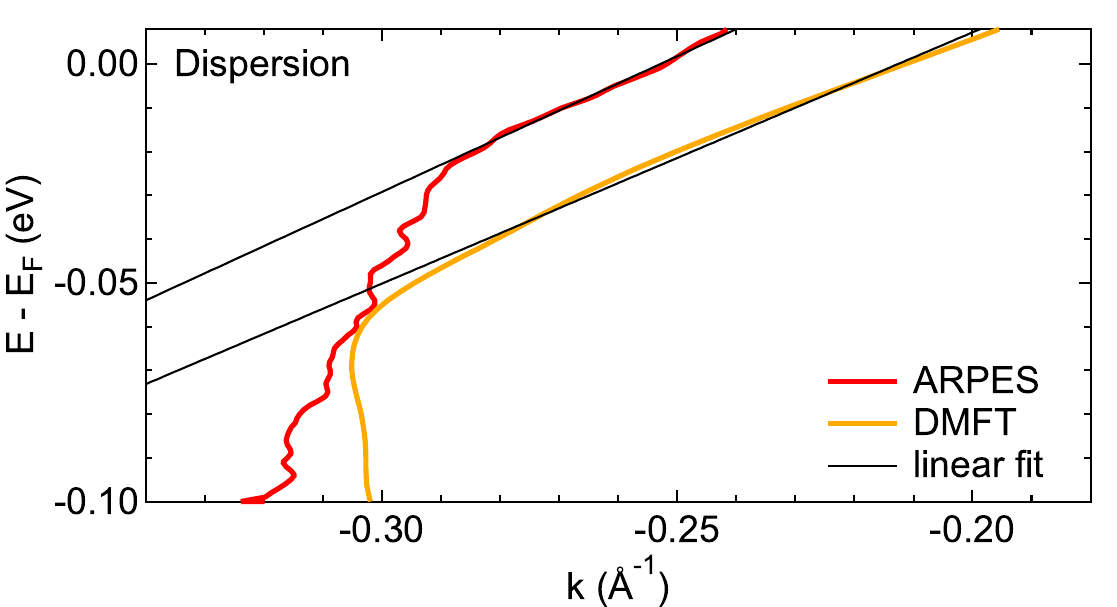}
\caption{
Fermi velocity. Low-energy section of the ARPES and DMFT dispersion shown in the main text in Fig.~2 are plotted together with linear fits around the Fermi energy.
}
\label{Fig:lowE_kink}
\end{figure}


\section{Supplementary Note 6: Photoemission Matrix Elements}

Photoemission dipole matrix elements $M = \langle f | \textbf{{A}}\cdot\textbf{{r}} | i \rangle$ were calculated in the length gauge using the approximation of a free electron final state and Fe$3d$ hydrogen-like wave functions as initial states \cite{goldberg_1978,gadzuk_1975,li_2024_matrix}. Results for the two experimental geometries are presented in Supplementary Fig.~\ref{Fig:matrix_elements}.

\begin{figure*}
\includegraphics[width={\textwidth}]{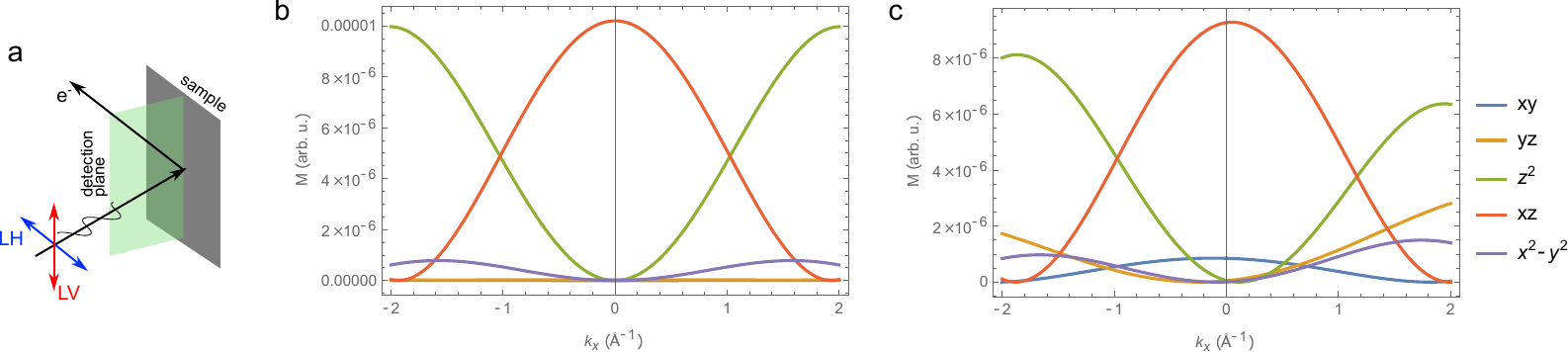}
\caption{
Photoemission matrix elements. {\bf{a}}: Experimental geometry. The sample is rotated with respect to the fixed beam and fixed analyzer in order to collect data along different momentum directions. {\bf{b}}: Orbital-resolved photoemission matrix elements for the geometry of Fig.~1 and 2 of the main text with $k_y=0$. {\bf{c}}: Same as {\bf{b}} but for Fig.~3 of the main text and accordingly $k_y=1.1$\,\AA$^{-1}$. The momentum axes match those used for the corresponding ARPES images in the main text.
}
\label{Fig:matrix_elements}
\end{figure*}


\bibliography{supplement}